\documentstyle[a4,12pt]{article}
 
\begin{document}
\titlepage
\begin{flushright}
LPTM 97/33\\
SPhT 97/85\\
hep-th/9707246 \\
July 1997
\end{flushright}
\vskip 1cm
\begin{center}
{\bf \Large
 Open Supermembranes Coupled to M-Theory}
\end{center}

\begin{center}
{\bf \Large 
Five-Branes}
\end{center}
\vskip 1cm
\begin{center}
{\large Ph. Brax$^{a}$\footnote{email: brax@spht.saclay.cea.fr} 
$\&$ J. Mourad$^{b}$\footnote{email: mourad@qcd.th.u-psud.fr}}
\end{center}
\vskip 0.5cm
\begin{center}
$^a$ {\it Service de Physique Th\'eorique, 
CEA-Saclay\\
F-91191 Gif/Yvette Cedex, France}\\
$^b$ {\it Laboratoire de Physique Th\'eorique
et Mod\'elisation, \\Universit\'e de Cergy-Pontoise,
Site Saint-Martin,\\ F-95302 Cergy-Pontoise, France}
\end{center}
\vskip 2cm
\begin{center}
{\large Abstract}
\end{center}

\noindent
We consider open supermembranes in eleven 
dimensions in the presence of
closed M-Theory five-branes. 
It has been  shown that, in a flat 
space-time, 
the world-volume action is kappa
invariant and preserves a fraction
of the eleven dimensional supersymmetries  if 
the boundaries of the membranes
lie on the 
five-branes. 
We calculate the reparametrisation anomalies due to the 
chiral fermions on the boundaries of the 
membrane and examine their cancellation mechanism. 
We show that these anomalies cancel
with the aid of a classical  term in the world-volume action,  
provided that the tensions of the 
five-brane and the membrane are related 
to the eleven dimensional gravitational
constant in a way already noticed in M-Theory. 
\newpage
\section{Introduction}
Since the recent advent of M-Theory as a step towards a 
unified description of the five previously known 
superstring theories, the role of extended objects 
generalising strings has been reappraised
\cite{rev}. 
The p-branes, whose origin 
stems from the corresponding (p+1)-forms and their duals
in the low energy spectrum of the various string theories, play a fundamental role in many non-perturbative phenomena and are 
 beleived to 
stand  on the same footing as the original string. 
The extended objects in the eleven dimensional 
M-Theory consist of a membrane coupled to a three-form in the spectrum
of the eleven dimensional  supergravity and a five-brane 
coupled to a six-form dual to the three-form. 
The dynamics of these six dimensional objects have 
recently received much attention \cite{f,f2,f3,f4,f5,f6,f7,f8,ff}, 
the non-linear Lagrangian whose double dimensional reduction 
leads to the type II-A 
D-four-brane action have been written in a 
covariant \cite{f6,f7} and non-manifestly covariant manner \cite{f8,ff}. 
A particular emphasis has been shed 
on the role of reparametrisation anomalies and their cancellation 
\cite{anf,anf2,anff}. 

The existence of closed supermembranes in eleven dimensions
 has been known for a few years \cite{b1,b2} leading by double 
dimensional reduction to the type II-A string 
in ten dimensions \cite{d1}. 
A similar construction for open supermembranes whose boundaries live on a 
ten  dimensional hypersurface was considered in \cite{bm}. 
It was shown that the only anomaly free 
configuration was the Horava-Witten one where the 
eleventh dimension is compactified on an 
orbifold $S^1/Z_2$ with an $E_8\times E_8$ gauge group 
living on the boundaries \cite{hw1,hw2}. 

In the present paper we further examine 
the configuration where the eleven dimensional 
supermembrane couples to five-branes 
via the strings living on its boundaries. This configuration has
received much attention recently especially
in relation with tensionless 
non-critical strings
in six dimensions \cite{te,te1,te2,te3,tee,se,ez}.
The goal of this paper is to calculate the reparametrisation anomaly localised
at the boundaries of the membrane and examine their cancellation mechanism.
In sections II and III we describe the 
dynamics of the supermembrane in the presence of
 five-branes. 
We shall use the non-covariant formulation 
of the five-brane action and show the influence of the boundaries.
A particular emphasis is put on the 
modification of the various Bianchi identities; 
the five-branes play the role magnetic 
charges for the M-Theory three-form
and the strings on the boundaries of the membranes 
are electric and magnetic sources for the 
five-brane self-dual two-form. 
In section IV we show that the Bianchi 
identities imply that the classical
action is not invariant under reparametrisation.
In section V we calculate the reparametrisation 
anomalies due to the chiral fermions on the strings. 
We show that the anomalies are given by the Euler class
of two SO(4) bundles: the normal bundle 
of the string in the five-brane and the 
normal bundle in eleven dimensions of a 
seven dimensional manifold containing both the membrane and the five-brane. 
We show how these anomalies 
are cancelled by the classical 
term in the world-volume action describing the coupling of the 
membrane and the five-brane
 without introducing new fields nor modifying the Bianchi identities.
The anomaly cancellation mechanism allows to 
rederive the relations between the membrane tension,
the five-brane tension and the eleven-dimensional 
gravitational constant. These relations 
have been previously obtained in different contexts \cite{anf,rel,rell,bm}.

\section{The Coupling Between a Supermembrane and a Five-Brane}

In reference \cite{bm} it was shown that the  kappa-invariant action of the 
open supermembrane in eleven dimensions
is given by
\begin{equation}
S=-T_3\left[\int_{\Sigma_3}\sqrt{- g}+
\int_{\Sigma_3} C -
\int_{\partial\Sigma_3} B\ \right],\label{a}
\end{equation}
where $ g_{ij}$ is the induced metric on the world-volume, 
$ C$ is the pullback of the eleven dimensional super three-form. We consider
a supermembrane which has the  topology of  
$\Sigma_2\times I$ where $I$ is the 
unit interval and $\Sigma_2$ a closed string. The boundaries of the
membrane correspond to two closed strings
\begin{equation}
\partial \Sigma_3=\Sigma_2^1\cup \Sigma_2^2.
\end{equation}
For the action (\ref{a}) to preserve a fraction of the supersymmetries while being kappa-invariant, the boundaries have to lie 
of an even-dimensional submanifold where
lives a two-form \cite{bm}. The two-form is essential to preserve the gauge invariance under
\begin{equation}
C\rightarrow C+d\Lambda,\label{g}
\end{equation}
with $\Lambda$ an arbitrary two-form.
The action (\ref{a}) is invariant under 
 (\ref{g}) if the two-form $B$ transforms as
\begin{equation}
B|_{\partial\Sigma_3}\rightarrow B|_{\partial \Sigma_3}-\Lambda|_{\partial\Sigma_3}.
\end{equation}
The five-brane, being six dimensional,
kappa-invariant, and having a two-form with the correct transformation rule under the 
gauge transformation (\ref{g})
is a natural candidate where the boundaries 
of the membrane can lie.
The interpretation of the eleven-dimensional five-brane as a Dirichlet
Brane for membranes was proposed previously in \cite{f}. 
The possibility that membranes can end on 
five-branes was pointed out in \cite{te1}
using charge conservation arguments. 

Since our membrane has two boundaries one has to distinguish between two cases, either the membrane has both of its boundaries on a single five brane (case A)
or the two boundaries of the membrane belong to two different parallel
five-branes (case B). In case (A) the membrane couples to a single
two-form and in case (B) it couples to two different two-forms. 
We shall find
it convenient to   orient the strings 
with the induced orientation from the five-branes so that in both cases (A) and (B) the boundary term in (\ref{a}) reads
\begin{equation}
I=\epsilon_1T_3\int_{\Sigma_2^1}B
+\epsilon_2T_3\int_{\Sigma_2^2}B,
\label{int}
\end{equation}
with $\epsilon_i=1$ if the induced orientations 
from the membrane and
the five-brane on $\Sigma_2^{i}$ coincide and $\epsilon_i=-1$ otherwise.
 
In the following section we examine the dynamics of the two-form $B$ and derive its Bianchi identity.

\section{The Five-Brane self-dual two-form action}

We consider a five-brane in eleven dimensions. 
The bosonic part of the five-brane contains a 
two-form $B$ with a self-dual field strength 
(in the linearised approximation). 
The corresponding supermultiplet is a $(2,0)$ tensor multiplet 
containing five scalars and two chiral spinors. 
We concentrate on the part containing the self-dual two-form in the linearised approximation.
The construction of the full kappa-invariant
 supercovariant action of the 
five-brane has been recently achieved \cite{f7}.
We use the non-covariant 
formulation \cite{f8} where the Lorentz invariance is not manifest.
The world-volume of the five-brane 
$\Sigma_6$ is chosen to be a circle bundle 
$\pi: \Sigma_6\to \Sigma_5$ with a base 
manifold $\Sigma_5$ and fibres $S^1$. 
Locally the first five coordinates $x^{\mu},\ \mu=0,\dots, 4$ 
parametrise the base space and $x_5$ the fibre.
 The local form of the self-dual two-form is
\begin{equation}
B=b+a\wedge dx_5
\end{equation}
where $b_{\mu\nu}$ is a two-form with indices $\mu,\nu=0\dots 4$
 and $a_{\mu}$ is a one-form.

The linearised action of $b$ is given by
\begin{equation}
S_6=-{T_6\over 4}\int_{\Sigma_6} h\wedge (*_6h-\partial_5b\wedge dx_5),\label{ac}
\end{equation}
where $*_6$ is the six dimensional Hodge dual map and
\begin{equation}
h=d_5b,
\end{equation}
with $d_5$ being the exterior derivative with respect to the coordinates of the base manifold $\Sigma_5$. The constant $T_6$ is the five-brane tension. 
Note that $*_6 h=*_5h\wedge dx_5$.
The equations of motion for $b$ read
\begin{equation}
d_5(*_5h-\partial_5b)=0.\label{e}
\end{equation}
From (\ref{e}) it follows that
there exits a one-form $a$ such that
\begin{equation}
*_5h=\partial_5b+d_5a
\end{equation}
In particular this implies that 
\begin{equation}
H=dB
\end{equation}
becomes
\begin{equation}
H=h+*_6h
\end{equation}
This implies that the action (\ref{ac}) describes a self-dual two-form.

Let us now consider topological defects coupled to the 
two-form $B$ of the five-brane. 
The action (\ref{ac}) has to be supplemented with
\begin{equation}
I=T_3\int_{\Sigma_2} B\vert_{\Sigma_2},
\end{equation}
where $\Sigma_2$ is the two-dimensional boundary of a three-manifold $\Sigma_3$ (a membrane). The coupling constant $T_3$ is the membrane tension. 
The submanifold $\Sigma_2$ is supposed to be embedded in $\Sigma_6$. One associates to the homology class $[\Sigma_2]$ a cohomology class $\delta(\Sigma_2,\Sigma_6)$ by Poincar\'e
duality such that
\begin{equation}
\int_{\Sigma_2}B\vert_{\Sigma_2}=\int_{\Sigma_6}B\wedge \delta(\Sigma_2,\Sigma_6).
\end{equation}
In the following, we find it convenient to choose $\Sigma_5$ such that
\begin{equation} 
\Sigma_2\subset\Sigma_5,
\end{equation}
so $I$ reads
\begin{equation}
T_3\int_{\Sigma_2}b,
\end{equation}
and we have
\begin{equation}
\delta(\Sigma_2,\Sigma_6)=\delta (\Sigma_2,\Sigma_5)\wedge dx_5.
\end{equation}
Using this four form 
the linearised equations of motion are 
\begin{equation}
d_5(*_5h-\partial b)={{2T_3}\over {T_6}}\delta (\Sigma_2,\Sigma_5).\label{dd}
\end{equation}
The solution of (\ref{dd}) reads
\begin{equation}
*_5h=\partial_5 b +d_5a +{2T_3\over T_6} \delta^1 (\Sigma_2,\Sigma_5).
\end{equation}
where we have used 
\begin{equation}
\delta (\Sigma_2,\Sigma_5)=d_5\delta^1(\Sigma_2,\Sigma_5),
\end{equation}
as $\Sigma_2$ is a closed surface. 
These equations of motion describe a self-dual two-form 
provided one modifies the definition of the field strength
\begin{equation}
H=dB+{2T_3\over T_6}\delta^1(\Sigma_2,\Sigma_6)
\end{equation}
This implies that 
\begin{equation}
H=h+*_6h,
\end{equation}
describing a self-dual field strength $H=*_6H$.
We now find that the Bianchi identity becomes 
\begin{equation}
dH={2T_3\over T_6} \delta(\Sigma_2,\Sigma_6)\label{bi}
\end{equation}
The string world-volume $\Sigma_2$ plays the role of an electric 
and  magnetic charge for the five-brane  
 that the Bianchi identity $dH=0$ is modified. 
In the presence of a background three-form $C$
the field strenght $H$ is modified as 
$H\rightarrow H-C|_{\Sigma_6}$, so the Bianchi identity
(\ref{bi}) becomes
\begin{equation}
dH={{2T_3}\over{T_6}}\delta(\Sigma_2,\Sigma_6)-G|_{\Sigma_6}.
\end{equation}
The Dirac quantisation condition applied in the six-dimensional five-brane 
gives
\begin{equation}
{{T_3^2}\over{T_6}}=m\pi,\label{d1}
\end{equation}
with $m$ an arbitrary integer. This relation will be useful in the following.

The presence of the five-brane modifies the Bianchi identity verified by the field
strenght $G$ of eleven dimensional supergravity. the Bianchi identity now reads
\begin{equation}
dG=-2\kappa_{11}^2T_6\delta(\Sigma_6,Q).
\end{equation}
Here and in the following $Q$ denotes the eleven-dimensional space-time and $\delta(\Sigma,\Sigma')$ with $\Sigma$ a d-dimensional submanifold of $\Sigma'$ is defined by
\begin{equation}
\int_{\Sigma}\omega=
\int_{\Sigma'}\omega\wedge\delta(\Sigma,\Sigma'),
\end{equation}
for an arbitrary form $\omega$ defined in $\Sigma'$. For future use we mention the relation
\begin{equation}
d\delta(\Sigma,\Sigma')=(-1)^{d+1}
\delta(\partial\Sigma,\Sigma').\label{dde}
\end{equation}
Similarly we shall denote by $N(\Sigma,\Sigma')$ the normal bundle of $\Sigma$ in $\Sigma'$. We shall also frequently
use a consequence of the Thom isomorphism theorem \cite{bt} which reads:
\begin{equation}
\delta(\Sigma,\Sigma')|_{\Sigma}=\chi\Big
(N(\Sigma,\Sigma')\Big),\label{ti}
\end{equation}
where $\chi$ is the Euler class of the normal bundle. 

Finally we mention that the Dirac quantisation condition in eleven dimensions 
gives the relation
\begin{equation}
\kappa_{11}^2T_6T_3=n\pi,\label{d2}
\end{equation}
with $n$ an arbitrary integer.
Note that the two relations (\ref{d1}) and 
(\ref{d2}) imply that there is only one physical scale in the theory.

\section{The classical transformation 
under reparametrisation}

We return to our configuration of a membrane having its two boundaries one one or on two five-branes. 
We examine separatly
the two cases (A) and (B) described in section 2.
Let us deal first with the case A where only one five-brane is present. 
The Bianchi identity for the field strengths $H$ and $G$ read
\begin{equation}
dG=-2\kappa_{11}^2T_6\delta (\Sigma_6,Q)
\label{bg}
\end{equation}
and
\begin{equation}
dH=2{T_3\over T_6}\Big(\epsilon_1 \delta (\Sigma_2^{1},\Sigma_6)+\epsilon_2
\delta(\Sigma_2^2,\Sigma_6)\Big) 
-G\vert_{\Sigma_6}.\label{bb}
\end{equation}
These two equations are compatible as
the differential of the right hand side of (\ref{bb}) vanishes. 
Indeed the strings on the boundary of the membrane are closed so the relation (\ref{dde}) gives $d\delta(\Sigma_2^{1,2},\Sigma_6)=0$
and
using the Thom isomorphism theorem  relating
\begin{equation}
\delta(\Sigma_6,Q)\vert_{\Sigma_6}=\chi(N(\Sigma_6,Q))
\end{equation}
and the fact that the Euler class of a manifold of odd dimension vanishes 
we get 
\begin{equation}
dG\vert_{\Sigma_6}=0
\end{equation}

With the aid of relation (\ref{dde})
one can integrate the Bianchi identity 
(\ref{bg}) to get
\begin{equation}
G\vert_{\Sigma_6}=-2\kappa_{11}^2T_6\delta (\Sigma_7,Q)+dC,
\end{equation}
where $\Sigma_7$ is a manifold whose boundary is $\Sigma_6$.
This leads to the three-form 
\footnote{ If another manifold $\Sigma_7'$ is chosen, this amounts to modifying $C'=C+\delta (\Sigma_8,Q)$ where the manifold
$\Sigma_8$ is such that $\partial \Sigma_8=\Sigma_7\cup \Sigma_7'$ glued on their common boundary $\Sigma_6$ with an opposite orientation, i.e the manifolds
$\Sigma_7$ and $\Sigma_7'$ are cobordant.}
\begin{equation}
H=dB-C\vert_{\Sigma_6}+2{T_3\over T_6}\Big(\epsilon_1\delta^1( \Sigma_2^1,\Sigma_6)
+\epsilon_2\delta^1( \Sigma_2^2,\Sigma_6)\Big)
+2\kappa_{11}^2T_6\delta^1(\Sigma_7,Q)\vert_{\Sigma_6}),
\end{equation}
where
\begin{equation}
\delta(\Sigma_7,Q)|_{\Sigma_6}=d\delta^1(\Sigma_7,Q).
\end{equation}

Let us now examine the variation of $B$ under a Lorentz gauge transformation.
Since $G$ is gauge invariant the gauge variation of $C$ is $ C\to C+d\Lambda$.
The action is invariant under the 
transformation (\ref{g}) so it is possible to choose $\delta C=0$.
 As $H$ is gauge invariant we get using
relation (\ref{ti})
\begin{equation}
\delta B\vert_{\Sigma_2^i}=-2\epsilon_i{T_3\over T_6}\chi^1_2(N_i)-2\kappa_{11}^2T_6\chi^1_2(N')\vert_{\Sigma_2^i},\ i=1,2,\label{tr}
\end{equation}
where $N_i=N(\Sigma_2^i,\Sigma_6)$,
$N'=N(\Sigma_7,Q)$,
and $\chi_2^1$ is related to  $\chi$ by
the descent equations: 
\begin{eqnarray}
\chi=d\chi_3,\nonumber \\
\delta\chi_3=d\chi_2^1.
\end{eqnarray}
The classical action is not invariant under the transformations
(\ref{tr}). Indeed,
the interaction term (\ref{int}) varies as
\begin{eqnarray}
\delta  I=-2{T_3^2\over T_6}\left(\int_{\Sigma_2^1}\chi_2^1(N_1)+
\int_{\Sigma_2^2}\chi_2^1(N_2)\right)\nonumber \\
-2\kappa_{11}T_6T_3\left(\epsilon_1
\int_{\Sigma_2^1}\chi_2^1(N')+\epsilon_2
\int_{\Sigma_2^2}\chi_2^1(N')\right).\label{vari}
\end{eqnarray}

The case B can be dealt with in a similar manner. The tensions of the
two-five branes, $\Sigma_6^1$ and $\Sigma_6^2$
 are denoted by $T_6^{(i)}, i=1,2$.
The Bianchi identities of 
the two three-forms $H_1$ and $H_2$ read
\begin{equation}
dH_i=2\epsilon_i{T_3\over T_6^{(i)}}\delta (\Sigma_2^i,\Sigma_6^i)
-G\vert_{\Sigma_6^i}.
\end{equation}
The five-branes are magnetic sources for the space-time three-form $C$, 
the corresponding Bianchi identity reads:
\begin{equation}
dG=-2\kappa_{11}^2\left(T_6^{(1)}\delta (\Sigma_6^1,Q)+T_6^{(2)}\delta (\Sigma_6^2,Q)\right).
\end{equation}
The gauge variation of the two-form $B_{i}$ then follows
\begin{equation}
\delta B_i\vert_{\Sigma_2^i}
=-2\epsilon_i{T_3\over T_6^{(i)}}\chi_2^1(N_i)-
2\kappa_{11}^2
T_6^{(i)}\chi_2^1(N'_i)\vert_{\Sigma_2^1}.
\end{equation}
The corresponding variation of the interaction term (\ref{int}) is then given by an expression which is similar to (\ref{vari}) and reads:
\begin{eqnarray}
\delta  I=-2{T_3^2
\over T_6^{(1)}}\int_{\Sigma_2^1}\chi_2^1(N_1)-2{T_3\over T_6^{(2)}}
\int_{\Sigma_2^2}\chi_2^1(N_2)\nonumber\\
-2\kappa_{11}T_3T_6^{(1)}\epsilon_1
\int_{\Sigma_2^1}\chi_2^1(N')-2\kappa_{11}T_3T_6^{(2)}
\epsilon_2
\int_{\Sigma_2^2}\chi_2^1(N').\label{varia}
\end{eqnarray}

In summary,
 the Bianchi identities imply that the classical 
world-volumw action is not invariant under reparametrisations.
 The variation of the action is given in the two cases involving one or
two
five-branes 
by equations (\ref{vari}) and (\ref{varia}) respectively.

\section{The Reparametrisation Anomaly }
The aim of this section is the calculation of the 
anomaly of the diffeomorphisms
that leave the membrane and the five-brane invariant. 
The closed membrane is anomaly-free so we  
concentrate on the anomalies that are 
localised on the intersection between the 
five-brane $\Sigma_6$ and the open membrane $\Sigma_3$.
The potential source of anomalies
are the Green-Schwarz fermions
which form a 32-components Majorana spinor of $SO(10,1)$.
The reparametrisation anomaly stems from the fermionic zero modes. 
Let us consider 
the case of a cylindrical membrane. The zero modes are independent
of the transverse direction to the boundary. As such the anomaly receives
contributions from the two strings on the boundary of the membrane.
This is similar to the anomaly calculation performed in \cite{hw2}.
In the following we calculate the anomaly due to the zero modes
in this cylindrical situation. These zero modes are the reduction of the
eleven dimensional spinor to the two dimensional surface, $\Sigma_2$
equipped with the orientation induced from the five-brane.

Let $N(\Sigma_6,Q)$ be the normal bundle 
to $\Sigma_6$ in $Q$, then
\begin{equation}
 TQ|_{\Sigma_2}=
T\Sigma_6|_{\Sigma_2}\oplus N(\Sigma_6,Q)|_{\Sigma_2}
\end{equation}
where $\Sigma_2$ denotes the the two topologically equivalent
boundaries of the membrane.
A  Lorentz transformation that leaves $T\Sigma_6$ invariant is equivalent to an
$SO(5,1)$ Lorentz rotation of $T\Sigma_6$
and an $SO(5)$ gauge transformation of 
$N(\Sigma_6,Q)$. 
The 32 complex spinors $SO(10,1)$ decompose as 
$(4_+,4)+(4_-,4)$ under $SO(5,1)\times SO(5)$. 
The kappa-symmetry condition  of the five brane eliminates the $(4_-,4)$
 component so the  fermions leading to the anomaly are 
Weyl fermions in the $(4_+,4)$ representation. 
The fact that the kappa-symmetry removes the fermions 
with negative chirality is due to our choice of the  
orientation of $\Sigma_2$ as the one induced from the five-brane.
 
Similarly let $N(\Sigma_3,Q)$ be the normal bundle to $\Sigma_3$ in $Q$ then 
\begin{equation}
TQ|_{\Sigma_2}=
T\Sigma_3|_{\Sigma_2}\oplus N(\Sigma_3,Q)|_{\Sigma_2}
\end{equation}
A  Lorentz transformation that leaves $T\Sigma_3$ invariant is equivalent to an
$SO(2,1)$ Lorentz rotation of $T\Sigma_3$
and an $SO(8)$ gauge transformation of 
$N(\Sigma_3,Q)$. The  32 of $SO(10,1)$ 
decomposes as $(2,8_+)+(2,8_-)$ under
$SO(2,1)\times SO(8)$. 
The component which is eliminated by the 
kappa-symmetry depends on the orientation of  $\Sigma_2$.
The kappa-symmetry condition
of the membrane eliminates the $(2,8_-)$ component 
if the induced orientations from the membrane and the five-brane coincide 
and we are left with the $(2,8_+)$ representation, 
otherwise  it is the $(2,8_+)$ component which is eliminated.  
We summarise these two cases by 
stating that it is $(2,8_{\epsilon})$ component  which remains.

We are interested in the diffeomorphisms thet leave both 
$T\Sigma_6$ and $T\Sigma_3$ invariant. They 
 leave $T\Sigma_2=T\Sigma_6\cap T\Sigma_3$ invariant.
Since $T\Sigma_6=T\Sigma_2\oplus N(\Sigma_2,\Sigma_6)$ and
$T\Sigma_3=T\Sigma_2\oplus N(\Sigma_2,\Sigma_3)$, the transformations of interest are
i) Lorentz rotation of $SO(1,1)$, ii) $SO(4)$
gauge transformations of
\begin{equation}
 N(\Sigma_2,Q)\cap T\Sigma_6\cap N(\Sigma_3,Q)=N(\Sigma_2,\Sigma_6)\equiv N
\end{equation}
and  iii) $SO(4)$ gauge transformations of 
\begin{equation}
N(\Sigma_2,Q)\cap N(\Sigma_6,Q)\cap N(\Sigma_3,Q)\equiv N'.
\end{equation}
 This is so because 
$ N(\Sigma_6,Q)\cap T\Sigma_3=N(\Sigma_6,Q)\cap N(\Sigma_2,\Sigma_3)$ is
of real rank one.
The bundle $N'$ is restriction to $\Sigma_2$ of the normal bundle
of a seven manifold $\Sigma_7$ such that 
\begin{equation}
\Sigma_6=\partial\Sigma_7,\ \Sigma_3\subset\Sigma_7
\end{equation}
and
\begin{equation}
N'=N(\Sigma_7,Q)\cap N(\Sigma_2,Q)
\end{equation}

We can now determine the representation 
of the physical Green-Schwarz fermion under
the $SO(1,1)\times SO(4)\times SO(4)$ resulting gauge group.
These are the representations that are in common from  the decomposition of
the $(4_+,4)$ of $SO(5,1)\times SO(5)$ and the $(2,8_{\epsilon})$ of $SO(2,1)\times SO(8)$ under the group $SO(1,1)\times SO(4)\times SO(4)$. Since $(4_+,4)=(1_+,2_+,4)+(1_-,2_-,4)$ and
$(2,8_{\epsilon})=(2,2_+,2_{\epsilon})+
(2,2_{-},2_{-\epsilon})$
the fermions are in the 
representation $(1_+,2_+,2_{\epsilon})+
(1_-,2_-,2_{-\epsilon})$
of $SO(1,1)\times SO(4)\times SO(4)$. The left and right handed fermions
transform under different representations of the gauge group $SO(4)\times
SO(4)$, so there are potential anomalies. 
The resulting anomaly reads
\begin{equation}
2\pi\left[ch\left(S_+(N)\otimes S_{\epsilon}(N')\right)
-ch\left(S_-(N)\otimes S_{-\epsilon}(N')\right)\right]
\hat{A}(T\Sigma_2),
\end{equation}
where $ch$ is the Chern character, $S_{\pm}$ is the spin bundle 
with a given chirality, $\hat A$ is the Dirac genus.
In order to calculate the Chern character
of an $SO(4)$  spin bundle  with a given chirality 
let $\lambda_1$ and $\lambda_2$ be the 
Chern roots of the curvature in the fundamental representation of $SO(4)$ then
\begin{equation}
ch(S_{\pm})=e^{(\lambda_1\pm\lambda_2)}+
 e^{-(\lambda_1\pm\lambda_2)}=
2+{{\lambda_1^2+\lambda_2^2\pm 2\lambda_1\lambda_2}\over{4}}+\dots.
\end{equation}
The first Pontryagin class is given by $p_1=\lambda_1^2+\lambda_2^2$ and the Euler 
class is given by $\chi=\lambda_1\lambda_2$ so we get 
\begin{equation}
ch(S_{\pm})=2+{{(p_1\pm 2\chi)}\over{4}}.
\end{equation}
Finally using the multiplicativity of the Chern character the total anomaly reads
\begin{equation}
I_{4}=4\pi( \chi(N)+\epsilon \chi(N')).
\end{equation}
We have kept the terms of degree four (the four forms) only. 
This anomaly has to be equally distributed between the two boundaries
of the membrane.
We find therefore that the total anomaly 
reads
\begin{equation}
\delta\Gamma=2\pi 
\int_{\Sigma_2^1}(\chi_2^1 (N_1)+\epsilon_1\chi_2^1 (N'))+2\pi
\int_{\Sigma_2^2}(\chi_2^1 (N_2)+\epsilon_2\chi_2^1 (N')). 
\end{equation}
It is valid in
both cases A and B.

This anomaly can be cancelled without introducing new fields.
In fact consider the total variation
due to the classical term $I$ and to the quantum anomaly,  it reads, in the case (A),
\begin{eqnarray}
\delta I +\delta\Gamma=
\left(2\pi-2{T_3^2 \over T_6}\right) 
\left(\int_{\Sigma_2^1}\chi_2^1 (N_1)+
\int_{\Sigma_2^2}\chi_2^1 (N_2)\right)
\nonumber\\
+
\left(2\pi-2\kappa_{11}^2T_3T_6\right)
\left(\epsilon_1\int_{\Sigma_2^1}\chi_2^1 (N')+\epsilon_2
\int_{\Sigma_2^2}\chi_2^1 (N')\right).
\end{eqnarray}
The variation cancels if
\begin{eqnarray}
T_3T_6\kappa_{11}^2&=&\pi\nonumber \\
T_3^2=\pi T_6.\label{rela}
\end{eqnarray}
This fixes the
integers apearing in the Dirac quantisation conditions (\ref{d1})
and (\ref{d2}) to one   
and is in agreement with
other derivations of these relations\cite{anf,rel,rell,bm}.

Case (B) can be dealt with in a similar manner where we get relations 
similar to (\ref{rela}) and in addition we get that 
the two five-brane tensions must be equal.

Notice that the anomaly cancellation mechanism is highly non-trivial
due to the structure of
$\delta \Gamma$ which is
the same as that of $\delta I$ and to the 
compatibility requirement with the Dirac conditions.

This
mechanism is very different from the 
one used for the Horava-Witten configuration
\cite{hw1,bm}
where extra fermions were required. 
These fermions are naturally coupled
to gauge fields. Here gauge fields are not present in the five-brane spectrum
preventing the possible use of extra fermions. With hindsight this reinforces
the result that the quantum consistency of the membrane-five-brane coupling   
follows from the modified Bianchi identities.

We can use the quantum consistency of the membrane coupling to a five-
brane in order to derive ten-dimensional 
configurations which are anomaly free. Suppose that
$x^0,\dots,x^5$ span the five-brane and $x^0,x^5,x^6$ span the membrane.
First of all choose one of the coordinates $x^7\in S^1_R$ where $R\to 0$,
then the M-theory reduces to the type II-A string. The five-brane reduces
to a type II-A NS solitonic five-brane and the membrane to a D-2-brane.
In the case A the D-2-brane is coupled to a single NS five-brane while in the
case B the D-2-brane connects two NS five-branes.
A second possibility is to take $x^4\in S^1_R,\ R\to 0$. The five-brane
reduces to a type II-A magnetic D-4-brane and the membrane to a D-2-brane.
A third possibility consists in choosing $x^5\in S_R^1,\ R\to 0$.
The five-brane reduces to a type II-A magnetic 4-brane and the
membrane to an open string. Finally, if $x^6\in S^1_R,\ R\to 0$ 
along the transverse direction of the membrane, then the five-brane
yields a type II-A solitonic NS five-brane and the membrane becomes a 
tensionless closed string.

\section{Conclusion}
We have analysed the coupling of the 
five-brane and the membrane of M-Theory and proved its quantum consistency. 
We have shown that the Bianchi identities for
the four-form of eleven dimensional supergravity and for
the self-dual three-form of the five-brane imply that
the classical world-volume action is not invariant 
under reparametrisation.
 The  boundaries of the membrane 
lie on the five brane where the Green-Schwarz 
fermions are chiral. 
These chiral fermions are the source of a reparametrisation 
anomaly. This anomaly is cancelled by the classical variation
provided that membrane tension, the five-brane tension and the 
eleven-dimensional gravitational constant are suitably related. 
It is remarkable that these relations are in agreement with other
approaches \cite{anf,rel,rell,bm}.
Together with the 
Horava-Witten configuration this gives 
a comprehensive picture of the role of open membranes in relation to
M-Theory.

\end{document}